\documentstyle[preprint,aps]{revtex}
\tightenlines

\newcommand{\nc}{\newcommand}
\nc{\be}{\begin{equation}}
\nc{\ee}{\end{equation}}
\nc{\bea}{\begin{eqnarray}}
\nc{\eea}{\end{eqnarray}}
\nc{\beas}{\begin{eqnarray*}}
\nc{\eeas}{\end{eqnarray*}}
\nc{\noi}{\noindent}
\nc{\sD}{\not \! \! D}
\nc{\s}[1]{\not \! #1}
\nc{\non}{\nonumber}
\nc{\bb}{\bibitem}
\nc{\lf}{\left}
\nc{\ri}{\right}
\nc{\mb}[1]{\makebox[#1]{}}
\nc{\pa}{\partial}
\nc{\sA}{\not \! \! A}
\nc{\newsec}[1]{\section{#1}\mb{0.5cm}}
\nc{\h}{\frac{1}{2}}
\nc{\ra}{\rightarrow}
\nc{\prw}{\widetilde{\Pi}_{\rho\omega}}
\nc{\la}{\leftarrow}
\nc{\etwopi}{$e^+e^-\ra\pi^+\pi^-\;$}
\nc{\ethrpi}{$e^+e^-\ra\pi^+\pi^0\pi^-\;$}
\nc{\lapp}{\hbox{$ {     \lower.40ex\hbox{$<$}
                   \atop \raise.20ex\hbox{$\sim$}
                   }     $}  }
\nc{\rapp}{\hbox{$ {     \lower.40ex\hbox{$>$}
                   \atop \raise.20ex\hbox{$\sim$}
                   }     $}  }
\nc{\M}{{\cal M}}
\nc{\rhoom}{$\rho^0$-$\omega\;$}
\nc{\rw}{$\rho$-$\omega\;$}
\def\mathunderaccent#1{\let\theaccent#1\mathpalette\putaccentunder}
\def\putaccentunder#1#2{\oalign{$#1#2$\crcr\hidewidth
\vbox to.2ex{\hbox{$#1\theaccent{}$}\vss}\hidewidth}}

\nc{\ti}{\mathunderaccent\tilde}

\begin{document}
\preprint{\vbox{                        \hfill UK/TP 98-05 \\
                                        \null\hfill hep-ph/9809224 \\
                                        \null\hfill September 1998\\
					Phys.Rev.D59:076002,1999}}

\title{Extracting Br$(\omega\ra\pi^+\pi^-)$ 
from the Time-like Pion Form-factor}

\author{S. Gardner\thanks{e-mail:
svg@ratina.pa.uky.edu} and H.B. 
O'Connell\thanks{e-mail: hoc@ruffian.pa.uky.edu}}
\address{Department of Physics and Astronomy,
University of Kentucky,\\
Lexington, KY 40506-0055 }
%\date{\today}
\maketitle

\begin{abstract}
We extract the G-parity-violating branching ratio 
Br$(\omega\ra\pi^+\pi^-)$ 
from the effective \rhoom mixing matrix element 
$\widetilde\Pi_{\rho\omega}(s)$, 
determined from \etwopi data. 
The $\omega\ra\pi^+\pi^-$ partial width can be determined either from the
time-like pion form factor or through
the constraint that the mixed physical
propagator $D_{\rho\omega}^{\mu\nu}(s)$ possesses no poles. 
The two procedures are inequivalent in
practice, and
we show why the first is preferred, to find finally 
Br$(\omega\ra \pi^+ \pi^-) = 1.9\pm 0.3\%$.

\end{abstract}
\pacs{PACS numbers: 11.30.Hv, 12.40.Vv, 13.25.Jx, 13.65.+i}

\narrowtext

\newpage

\section{Introduction}

The presence of the $\omega$
resonance in \etwopi, in the region
dominated by the $\rho^0$, signals the presence of
the G-parity-violating decay 
$\omega\ra\pi^+\pi^-$. 
Our purpose is to extract the value of 
Br$(\omega\ra\pi^+\pi^-)$ from fits to \etwopi in the 
\rhoom interference region. To do so, we must consider the
relationship between the partial width 
$\Gamma(\omega \ra \pi^+\pi^-)$ and the 
effective \rhoom mixing matrix element 
$\widetilde\Pi_{\rho\omega}(m_\omega^2)$, 
determined in our earlier 
fits~\cite{GO} to \etwopi data~\cite{Bark,wd85}. 
The \etwopi cross section $\sigma(s)$
can be written as 
$\sigma(s)=\sigma_{\rm em}(s)|F_\pi(s)|^2$, where $\sigma_{\rm em}(s)$
is the cross section for the production of a structureless $\pi^+\pi^-$
pair and $s$ is the usual Mandlestam variable. 
The time-like pion
form factor $F_\pi(s)$ can in turn be written, to leading order
in isospin violation, as~\cite{GO}
\be
F_\pi(s)=F_\rho(s)\left[1+
\frac{1}{3}
\left(\frac{\widetilde{\Pi}_{\rho\omega}(s)}{s-m_\omega^2+
i m_\omega\Gamma_\omega}\right)\right]\;, 
\label{pionff}
\ee
where $F_\rho(s)$ parametrizes the $\rho^0$ resonance and 
$\widetilde\Pi_{\rho\omega}(s)$ is the effective
\rhoom mixing matrix element noted earlier. 
$\Gamma(\omega\ra\pi^+\pi^-)$ is determined by the effective
$\omega\ra\pi^+\pi^-$ coupling constant $g_{\omega\pi\pi}^{\rm eff}$, 
which can be 
extracted either from the
time-like pion form factor or from 
the relationship between the physical and isospin-perfect 
vector meson fields, determined through
the constraint that the mixed physical
propagator $D_{\rho\omega}^{\mu\nu}(s)$ possesses no poles. 
We evaluate not only the relationship
between these two different methods 
but also the impact of the 
uncertainty in the $\rho^0$ mass and width on Br$(\omega\ra\pi^+\pi^-)$
before reporting our final results. Despite the close connection 
between Br$(\omega\ra\pi^+\pi^-)$ and $\widetilde\Pi_{\rho\omega}(s)$,
we believe this work represents the first attempt to determine 
both simultaneously from \etwopi data.

\section{$\Gamma(\omega\ra\pi^+\pi^-)$ and 
\rhoom Mixing}

If isospin symmetry were perfect, the $\rho$ and $\omega$ resonances would 
be exact eigenstates of G-parity, so that the $\rho$, of even 
G-parity, would 
decay to two, but not three, 
pions and the $\omega$, of odd G-parity, would decay 
to three, but not two, pions. 
Yet this is not strictly so, for \rhoom
 interference in \etwopi is
observed in nature~\cite{miller90}. Nevertheless, it 
is useful to introduce an isospin-perfect basis $\rho_I^0$ and 
$\omega_I$ in which to describe the physical $\rho^0$ and $\omega$. 
In this basis, 
G-parity can be violated either through ``mixing'', 
$\langle\omega_I| H^{\rm mix} |\rho_I\rangle$, where
$H^{\rm mix}$ represents isospin-violating terms in the effective
Hamiltonian in the vector meson sector,
or through the
direct decay $\langle\omega_I| H^{\rm mix} | \pi^+\pi^-\rangle$.
The vector mesons in \etwopi couple to 
a conserved current, so that 
we can write their propagators as 
$D_{VV}^{\mu\nu}(s)
\equiv g^{\mu\nu}D_{VV}(s)$, thereby defining
the scalar part of the propagator, $D_{VV}(s)$. The propagator
possesses a pole in the complex plane at $s=z_V$, so that
in the vicinity of this pole we have $D_{VV}(s)=1/(s-z_V)\equiv1/s_V$.
The difference between the diagonal scalar propagator 
in the physical and 
isospin-perfect bases, i.e., between  
$D_{VV}(s)$ and $D_{VV}^I(s)$, 
 is of non-leading-order in isospin violation, so that 
$D_{VV}^I(s)=1/s_V$ as well. 
Consequently, the pion form factor in the resonance region 
in the isospin-perfect basis can be 
written, to leading order in isospin violation, as 
\be
F_\pi(s)=\frac{g_{\rho_I\pi\pi}f_{\rho_I\gamma}}{s_\rho}+
\frac{g_{\omega_I\pi\pi}f_{\omega_I\gamma}}{s_\omega}+
\frac{g_{\rho_I\pi\pi}\Pi_{\rho\omega}^{I}(s)
f_{\omega_I\gamma}}{s_\rho s_\omega} \;,
\label{ff-iso}
\ee
where $g_{V_I\pi\pi}$ and $f_{V_I\gamma}$ are the 
vector-meson--pion-pion and vector-meson--photon 
coupling constants, respectively. The first term reflects the
dominant process $\gamma\ra\rho^0\ra\pi^+\pi^-$, whereas the
G-parity-violating terms reflect 
the direct decay $\omega\ra\pi^+\pi^-$
and \rhoom mixing, $\omega\ra\rho^0\ra\pi^+\pi^-$, respectively, 
noting
the mixing matrix element $\Pi_{\rho\omega}^{I}(s)$.
Defining $G\equiv g_{\omega_I\pi\pi}/g_{\rho_I\pi\pi}$ we can rewrite
Eq.~(\ref{ff-iso}) as 
\bea
\non
F_\pi(s)&=&
\frac{g_{\rho_I\pi\pi} f_{\rho_I\gamma}}{s_\rho} + 
\frac{g_{\rho_I\pi\pi} f_{\omega_I\gamma}}{s_\rho s_\omega}
(G(s-z_\rho) + \Pi_{\rho\omega}^{I}(s)) \\
\qquad &\equiv&
\frac{f_{\rho_I\gamma} g_{\rho_I\pi\pi}}{s_\rho}
\left[1+\frac{f_{\omega_I\gamma}}{f_{\rho_I\gamma}}
\left(
\frac{\widetilde{\Pi}_{\rho\omega}(s)}{s-z_\omega}
\right)
\right] \;.
\label{eq:formnew}
\eea
Note that we have defined the effective mixing matrix element 
%!
$\widetilde{\Pi}_{\rho\omega}(s)$, as $G$ and 
$\Pi_{\rho\omega}^{I}(s)$ cannot be 
meaningfully separated in a fit to data~\cite{MOW,OTW}, 
for both terms are $s$-dependent~\cite{GHT,OPTW}. 
As $\Gamma_\omega \ll m_\omega$
a Breit-Wigner lineshape may be used to model the $\omega$
resonance, but the large width of the $\rho$ relative to
its mass obliges a more sophisticated treatment. Rather than
adopting $s_\rho=s - z_\rho$, appropriate for $s\approx z_\rho$,
for the entire resonance region,
we replace 
$f_{\rho_I\gamma} g_{\rho_I\pi\pi}/{s_\rho}$ by 
$F_\rho(s)$, a function constructed 
to incorporate 
the constraints imposed on the form factor by time-reversal
invariance, unitarity, analyticity, and 
charge conservation. For further details, 
see Ref.~\cite{GO}
and references therein. Using
\be
F_\pi(s)=F_\rho(s)\left[1+ \frac{f_{\omega_I\gamma}}{f_{\rho_I\gamma}}
\left(
\frac{\widetilde{\Pi}_{\rho\omega}(s)}{s-m_\omega^2+
im_\omega\Gamma_\omega}
\right)\right] 
\label{ff-isofit}
\ee
with the SU(6) value of $f_{\omega_I\gamma}/f_{\rho_I\gamma}=1/3$,
we find
$\widetilde{\Pi}_{\rho\omega}(m_\omega^2)=-3500\pm 300$ MeV$^2$, where
the systematic error due to the $\rho^0$ parametrization adopted
is negligible~\cite{GO}. 
Note that both the imaginary part of 
$\widetilde{\Pi}_{\rho\omega}(s)$, 
${\rm Im}\widetilde \Pi_{\rho\omega}(m_\omega^2) = - 300 \pm 300$
MeV$^2$, and its 
$s$-dependence about $s=m_\omega^2$,
$\widetilde{\Pi}_{\rho\omega}(s) = \widetilde{\Pi}_{\rho\omega}(m_\omega^2) 
+ (s - m_\omega^2)\widetilde{\Pi}'_{\rho\omega}(m_\omega^2)$
%!
with $\widetilde{\Pi}'_{\rho\omega}(m_\omega^2)=0.03 \pm 0.04$,
are also negligible~\cite{GO}.

Equation (\ref{eq:formnew}) can also
be used to define an effective, 
isospin-violating coupling constant, 
$g_{\omega\pi\pi}^{\rm eff}(s)$, such that 
\be
F_\pi(s)=
\frac{g_{\rho_I\pi\pi}f_{\rho_I\gamma} }{s_\rho} + 
\frac{g_{\omega\pi\pi}^{\rm eff}(s) f_{\omega_I\gamma}}{s_\omega} \;,
\label{ff-geff1}
\ee
so that 
$g_{\omega\pi\pi}^{\rm eff}(s)\equiv g_{\rho_I\pi\pi}
\widetilde{\Pi}_{\rho\omega}(s)/s_\rho$. 
To determine the partial width 
$\Gamma(\omega\ra\pi^+\pi^-)$, and hence 
Br$(\omega\ra\pi^+\pi^-)$, we must relate it to the 
effective coupling constant $g_{\omega\pi\pi}^{\rm{eff}}(s)$. 

In a Lagrangian model in which the pion is an elementary field and 
$g_{V\pi\pi}$ denotes the vector meson coupling constant to two
pions, the vector meson self-energy $\Pi_{VV}(s)$, noting 
$D_{VV}^{-1}(s)= s - m^2 - \Pi_{VV}(s)$, can be approximated
 as a sum of
iterated bubble diagrams, where each bubble contains a 
two-pion intermediate state~\cite{decay}. 
Here $g_{V\pi\pi}$ is a simple constant,
and direct calculation yields~\cite{decay}
\be
{\rm Im}\, \Pi_{VV}(s) = - g_{V\pi\pi}^2 
\frac{(s - 4m_\pi^2)^{3/2}}{48\pi \sqrt{s}} \Theta(s - 4m_\pi^2) \;.
\ee
Finally, noting $\lim_{s\ra m_V^2} {\rm Im}\, \Pi_{VV}(s)
= - m_V \Gamma (V\ra\pi^+\pi^-)$, then~\cite{decay,klingl}
\be
\Gamma(V\ra \pi^+\pi^-)=\frac{g^2_{V\pi\pi}}{48\pi}
\frac{(m_V^2-4m_\pi^2)^{3/2}}{m_V^2} \;.
\label{eq:gpp}
\ee
Replacing $g_{\omega\pi\pi}$ by 
$|g_{\omega\pi\pi}^{\rm{eff}}(s)|$, one finds that 
Br($\omega\ra\pi^+\pi^-$), to leading order in isospin
violation, is given by 
\be
{\rm Br}(\omega\ra\pi^+\pi^-)
=\frac{1}{48\pi}
\frac{(m_\omega^2-4m_\pi^2)^{3/2}}{m_\omega^2\Gamma_\omega}
\left| \frac{g_{\rho_I\pi\pi}\widetilde{\Pi}_{\rho\omega}(s)}
{s_\rho}
\right|^2 \Bigg|_{s=m_\omega^2} \;.
\label{Br1}
\ee

Another relation for $g_{\omega\pi\pi}^{\rm{eff}}(s)$ emerges
through consideration of the pion form factor in the 
physical basis \cite{basis1}. 
To leading order in isospin violation, 
we have~\cite{MOW,OTW}
\be
F_\pi(s) = g_{\rho\pi\pi} D_{\rho\rho} f_{\rho\gamma}
+ g_{\rho\pi\pi} D_{\rho\omega} f_{\omega\gamma}
 + g_{\omega\pi\pi} D_{\omega\omega} f_{\omega\gamma} \;,
\label{ff-phys}
\ee
where 
we introduce a \rhoom mixing  matrix element, $\Pi_{\rho\omega}(s)$,
such that~\cite{OPTW}
\be
D_{\rho\omega}^I(s)
=D^I_{\omega\omega}(s)\Pi_{\rho\omega}(s)D^I_{\rho\rho}(s).
\ee 
To relate the physical states $\rho$ and $\omega$ 
to the
isospin perfect ones $\rho_I$ and $\omega_I$, we introduce 
two mixing parameters,
$\epsilon_1$ and $\epsilon_2$, such that~\cite{MOW,OTW}
\be
\rho= \rho_I - \epsilon_1 \omega_I \quad ; \quad 
\omega= \epsilon_2 \rho_I + \omega_I \;.
\label{physdef}
\ee
Requiring the
mixed physical propagator 
$D_{\rho\omega}(s)$ to possess no poles, 
$\epsilon_1$ and $\epsilon_2$
are determined to be~\cite{MOW,OTW}:
\be
\epsilon_1=
\frac{\Pi_{\rho\omega}(z_\omega)}{z_\omega-z_\rho}
\,\,\,\,\,\,\,\,\,
\epsilon_2=
\frac{\Pi_{\rho\omega}(z_\rho)}{z_\omega-z_\rho} \;.
\ee
Using Eqs.~(\ref{ff-phys}) and (\ref{physdef}), 
and $D_{VV}=D_{VV}^I=1/s_V$ for $s$ in the vicinity of 
$z_\rho,z_\omega$ yields
\bea
\non
F_\pi(s)&=&\frac{g_{\rho_I\pi\pi}f_{\rho_I\gamma}}{s_\rho}
+
\frac{g_{\omega_I\pi\pi}}{s_\omega s_\rho} \\
&\times&
\left[
G(s - z_\rho) +  
\frac{(\Pi_{\rho\omega}(z_\rho) - \Pi_{\rho\omega}(z_\omega))}
{z_\omega - z_\rho} (s - z_\omega + s - z_\rho) + 
\Pi_{\rho\omega}(s)
\right]f_{\omega_I\gamma} \;. 
\label{ff-phys/iso}
\eea
Comparison with Eq.~(\ref{ff-iso}) shows that 
$\Pi_{\rho\omega}^I(s)$ and $\Pi_{\rho\omega}(s)$ are only
equivalent if $\Pi_{\rho\omega}(z_\rho) = \Pi_{\rho\omega}(z_\omega)$. 
We can then write $g_{\omega\pi\pi}^{\rm eff}$ found above 
as 
\bea
\non
g_{\omega\pi\pi}^{\rm eff}(s)
&=& \frac{g_{\rho_I\pi\pi}}{s_\rho}
\widetilde{\Pi}_{\rho\omega}(s) \\ 
&=& \frac{g_{\rho_I\pi\pi}}{s_\rho}
\left[
G(s - z_\rho) + 
\frac{(\Pi_{\rho\omega}(z_\rho) - \Pi_{\rho\omega}(z_\omega))}
{z_\omega - z_\rho} (s - z_\omega + s - z_\rho) + 
\Pi_{\rho\omega}(s)
\right] \;.
\label{geff1}
\eea
We could also have defined $g_{\omega\pi\pi}^{\rm eff}$ directly
from the relation between the physical and isospin perfect bases,
Eq.~(\ref{physdef}):
\bea
\non
g_{\omega\pi\pi}^{\rm eff}&=&
g_{\omega_I\pi\pi} + \epsilon_2 g_{\rho_I\pi\pi} \\
&=&
\frac{g_{\rho_I\pi\pi}}{z_\omega - z_\rho}
\left[
G(z_\omega - z_\rho) +
\Pi_{\rho\omega}(z_\rho)
\right] \;.
\label{geff2}
\eea
These two possible definitions of 
$g_{\omega\pi\pi}^{\rm eff}$ are {\it identical} at the 
$\omega$ pole, $s=z_\omega$. However, fits to the time-like
pion form factor data yield 
$\widetilde{\Pi}_{\rho\omega}(s)$ merely at real values of $s$, so
that Eq.~(\ref{geff1}) is the only practicable definition of
$g_{\omega\pi\pi}^{\rm eff}$. 
The two expressions differ in general as 
isospin-violating pieces are present in $f_{\rho\gamma}$ as well; 
they vanish, however, at $s=z_\omega$.

Interestingly, if we were to demand as in Ref.~\cite{OTW} that 
$\Pi_{\rho\omega}^I(s)\equiv \Pi_{\rho\omega}(s)$, implying that 
Eq.~(\ref{physdef}) cannot be used to relate 
$f_{V\gamma}$ to $f_{V_I\gamma}$ and
$g_{V\pi\pi}$ to $g_{V_I\pi\pi}$ unless $\epsilon_1=\epsilon_2$~\cite{OTW}, 
then Eq.~(\ref{geff1}) would become 
$g_{\omega\pi\pi}^{\rm eff}(s) = g_{\rho_I\pi\pi}
(G(s - z_\rho) + \Pi_{\rho\omega}(s))/s_\rho$. This latter
definition of $g_{\omega\pi\pi}^{\rm eff}(s)$ would be inconsistent
with Eq.~(\ref{geff2}) at $s=z_\omega$. We prefer the analysis 
yielding Eq.~(\ref{geff1}). 

To determine Br$(\omega\ra\pi^+\pi^-)$ using Eq.~(\ref{Br1}) 
we must evaluate $g_{\rho_I\pi\pi}/s_\rho$ at $s=m_\omega^2$. 
As $s_\rho=s-z_\rho$ only for $s\approx z_\rho$, it is 
appropriate to replace 
$g_{\rho_I\pi\pi}/s_\rho$
by $F_\rho(s)/f_{\rho_I\gamma}$, noting
Eqs.~(\ref{eq:formnew}) and (\ref{ff-isofit}), 
to yield finally 
\be
{\rm Br}(\omega\ra\pi^+\pi^-)
=\frac{(m_\omega^2-4m_\pi^2)^{3/2}}{48\pi
m_\omega^2\Gamma_\omega f_{\omega_I\gamma}^2}
\left|F_\rho(m_\omega^2)\;
\frac{1}{3}\;\widetilde{\Pi}_{\rho\omega}(m_\omega^2)\right|^2 \;.
\label{Br_final}
\ee
In the fit to data using Eq.~(\ref{ff-isofit}), 
$(f_{\omega_I\gamma}/f_{\rho_I\gamma})\widetilde{\Pi}_{\rho\omega}(s)$
appears as a single fitting parameter. Choosing 
$f_{\omega_I\gamma}/f_{\rho_I\gamma}=1/3$, then, allows us to 
use our earlier value of 
$\widetilde{\Pi}_{\rho\omega}=-3500$ MeV$^2$~\cite{GO}. 
Equation (\ref{Br_final}) defines the branching ratio in terms of the 
phenomenologically well-constrained
fitting functions $F_\rho(s)$ 
%!
and $\prw/3$ and thus avoids the explicit introduction of
$\rho$ resonance parameters. The model dependence 
of Eq.~(\ref{Br_final}) is therefore minimal, and 
for this reason it is our  preferred definition. 

%!
To assess its utility we shall compare it with other definitions
in the literature. We may also 
use Eq.~(\ref{eq:gpp}) to replace $g_{\rho_I\pi\pi}$ and 
write $s_\rho= s - m_\rho^2 + i m_\rho \Gamma_\rho$ to find
\be
{\rm Br}^{(2)}(\omega\ra\pi^+\pi^-)=
\frac{m_\rho^2(m_\omega^2-4m_\pi^2)^{3/2}}{m_\omega^2(m_\rho^2-4m_\pi^2)^{3/2}}
\frac{\Gamma_\rho}{\Gamma_\omega}
\left|\frac{\prw(m_\omega^2) }
{m_\omega^2 - m_\rho^2+im_\rho\Gamma_\rho}
\right|^2 \;,
\label{Br2}
\ee
where we have used $\Gamma_\rho=\Gamma(\rho\ra\pi^+\pi^-)$.
If we set $m_\omega=m_\rho$, 
Eq.~(\ref{Br2}) becomes that used in Ref.~\cite{CB} to extract 
$\prw(m_\omega^2)=-4520$ MeV$^2$~\cite{newCB}
a value commonly used in the literature~\cite{quoted}.
We prefer determining both $\prw(m_\omega^2)$ and 
Br($\omega\ra\pi^+\pi^-$) 
directly from our fits to 
the \etwopi data. 
Yet another expression for Br($\omega\ra\pi^+\pi^-)$ results
if we consider Eq.~(\ref{geff2}) in place of Eq.~(\ref{geff1}) for 
$g_{\omega\pi\pi}^{\rm eff}$; that is, 
\be
{\rm Br}^{(3)}(\omega\ra\pi^+\pi^-)=
\frac{m_\rho^2(m_\omega^2-4m_\pi^2)^{3/2}}{m_\omega^2(m_\rho^2-4m_\pi^2)^{3/2}}
\frac{\Gamma_\rho}{\Gamma_\omega}
\left|\frac{\overline{\Pi}_{\rho\omega}(m_\omega^2)}
{z_\omega - z_\rho}
\right|^2 \;,
 \label{Br3}
\ee
where $\overline{\Pi}_{\rho\omega}(m_\omega^2)\equiv 
G(z_\omega - z_\rho) + \Pi_{\rho\omega}(z_\rho)$. 
$\overline{\Pi}_{\rho\omega}(m_\omega^2)$ is not determined directly
in fits to \etwopi data and thus we favor Eqs.~(\ref{Br_final}) or
(\ref{Br2}). Nevertheless, 
as we found no significant $s$-dependence to
$\prw$ 
in our fits to
\etwopi data~\cite{GO}, we will replace 
$\overline{\Pi}_{\rho\omega}(m_\omega^2)$
by $\widetilde{\Pi}_{\rho\omega}(m_\omega^2)$ in our
subsequent numerical estimates. 
Neglecting terms of ${\cal O}((m_\omega-m_\rho)/m^{\rm av})$, with 
$m^{\rm av}=(m_\rho + m_\omega)/2$, and setting 
$z_\rho=m_\rho^2 + i m_\rho\Gamma_\rho$,
Eq.~(\ref{Br3}) yields
\be
\Gamma(\omega\ra\pi^+ \pi^-) = 
\frac{\left|\overline{\Pi}_{\rho\omega}(m_\omega^2)\right|^2}
{4 m_\rho^2((m_\omega - m_\rho)^2 + 
{1\over 4} (\Gamma_\omega - \Gamma_\rho)^2)}\Gamma_\rho \;,
\ee
and is thus equivalent to Eq.(B.12) in Ref.~\cite{GL}.
So far we have freely changed from one realization of $s_\rho$
to another; i.e., we have written both $s_\rho=s - z_\rho$ and
$s_\rho=s- m_\rho^2 + i m_\rho \Gamma_\rho$. Yet it is important
to recognize that for a broad resonance, such as the $\rho$ (but 
unlike the $\omega$), these
realizations are not necessarily equivalent. A parametrization of 
$F_\rho(s)$ which explicitly suits the constraint of 
unitarity and time-reversal
invariance, obliging 
its phase to be that of $l=1$, $I=1$ $\pi$-$\pi$ scattering for $s$ where
the scattering is elastic~\cite{gas66,HL,GO}, 
results in an $s$-dependent width~\cite{GS}. Effectively, then,
$(F_\rho(s))^{-1}\propto s - m_\rho^2 + i m_\rho \Gamma_\rho(s)$, where
the $m_\rho$ and $\Gamma_\rho$
we have used thus far satisfy $\Gamma_\rho\equiv \Gamma_\rho(m_\rho^2)$.
However,
the $\rho$ pole, $z_\rho$, in the complex $s$ plane is 
determined by requiring $(F_\rho(z_\rho))^{-1}=0$. 
Thus, in the presence of a $s$-dependent width, $\Gamma_\rho(s)$, 
$z_\rho \ne m_\rho^2 - im_\rho \Gamma_\rho$. 
If we parametrize $z_\rho$ as 
\be
z_\rho\equiv\overline{m}^2_\rho-i\overline{m}_\rho\overline{\Gamma}_\rho\;,
\label{pole}
\ee
then 
$\overline{m}_\rho$ and $\overline{\Gamma}_\rho$ differ substantially
from $m_\rho$ and $\Gamma_\rho$~\cite{lang79}, as illustrated in 
Table \ref{one}. Moreover, 
$\overline{m}_\rho$ and $\overline{\Gamma}_\rho$ 
are independent
of the parametrization of $F_\rho(s)$~\cite{Levy,Smatrix,pdg72,lang79}, 
whereas
$m_\rho$ and $\Gamma_\rho$ are {\it not}~\cite{pisut68,ben93,PDG,GO}.
In marked contrast to $m_\rho$ and
$\Gamma_\rho$ given in Table~\ref{one},
the average values of $\overline{m}_\rho$
and  $\overline{\Gamma}_\rho$, 
\be
\overline{m}_\rho=757.0 \pm 1.1\,{\rm MeV}\;,\,\,\,\,\,\,\,\,\,\,\,\,
\overline{\Gamma}_\rho=141.3 \pm 3.1\,{\rm MeV} \;,
\ee
are within one standard deviation of the 
$\overline{m}_\rho$ and $\overline{\Gamma}_\rho$ found in 
each and every model. 
This is in excellent agreement with Ref.~\cite{BCP}, where the
$\rho$ parameters are found to be 
$\overline{m}_\rho=757.5\pm1.5$ MeV and 
$\overline{\Gamma}_\rho=142.5\pm3.5$ MeV. 
The stability shown here is that of the
S-matrix
pole position, $z_\rho$, which is 
model independent~\cite{Levy,Smatrix,pdg72,lang79}.
The separation of $z_\rho$ into a ``mass'' and
``width'', as in Eq.~(\ref{pole}),
though useful \cite{Levy}, 
is somewhat artificial, as Re($\sqrt{z_\rho}$) and
$\sqrt{{\rm Re}\,z_\rho}$ could equally well serve as 
the mass~\cite{Stuart}. 
We shall consider the consequence of 
$z_\rho \ne m_\rho^2 - im_\rho \Gamma_\rho$ on the numerical values of
${\rm Br}^{(3)}(\omega\ra\pi^+\pi^-)$. 

It should also be noted that 
the value of 
$\widetilde{\Pi}_{\rho\omega}(m_\omega^2)$
to be used in Eqs.~(\ref{Br2}) and (\ref{Br3}) 
can be determined 
from our previous, averaged 
result~\cite{GO}, noting Eq.~(\ref{ff-isofit}),
through
\be
\widetilde{\Pi}_{\rho\omega}(m_\omega^2)=
\frac{1}{3}\frac{f_{\rho_I\gamma}}{f_{\omega_I\gamma}}
\left( -3500\, {\rm MeV}^2
\right) \;, 
\label{replace}
\ee
We must therefore now determine the leptonic couplings 
$f_{\rho_I\gamma}$ and $f_{\omega_I\gamma}$.

\section{Vector meson electromagnetic couplings}
We have related the branching ratio Br$(\omega\ra\pi^+\pi^-)$
to the effective mixing term $\prw(s)$ and 
various vector-meson parameters, yet in order 
to fix $\prw$ in a 
fit to \etwopi data, 
we need to determine the ratio $r_\gamma\equiv f_{\rho_I\gamma}/
f_{\omega_I\gamma}$. 
In the SU(6) limit
$r_\gamma=3$, but
this relation is broken
at the $\sim 10\%$ level \cite{klingl} by the large $\rho$ width 
\cite{GS,renard}. 
In this section we discuss the
extraction of $f_{\rho_I\gamma}$ and $f_{\omega_I\gamma}$.

The vector-meson--photon coupling constant
$f_{V\gamma}$ is
related to the leptonic decay width $\Gamma(V\ra \ell^+\ell^-)$ through 
\be
\Gamma(V\ra \ell^+\ell^-)
=\frac{4\pi\alpha^2}{3m_V^3}f^2_{V\gamma } \;,
\label{eq:lep}
\ee
noting
that lepton masses enter at ${\cal O}((m_\ell/m_V)^4)$~\cite{klingl}.

The cross-section for \etwopi, proceeding solely through 
$e^+e^-\ra \rho^0 \ra \pi^+\pi^-$, that is, assuming no background,
for $s=m_\rho^2$ is 
\bea\non
\sigma(e^+e^-\ra\rho_I\ra\pi^+\pi^-) 
&=&\left.\frac{\pi\alpha^2}{3}\frac{(s-4m_\pi^2)^{3/2}}{s^{5/2}}
\frac{\left(f_{\rho_I\gamma}g_{\rho_I\pi\pi}\right)^2}{(s-m_\rho^2)^2
+m_\rho^2\Gamma^2_\rho}\right|_{s=m_\rho^2} \\
&=&12\pi\frac{\Gamma(\rho_I\ra e^+e^-)\Gamma(\rho_I\ra \pi^+\pi^-)}
{m_\rho^2\Gamma_\rho^2} \;,
\label{picase}
\eea
where we have used Eqs.~(\ref{eq:gpp}) and (\ref{eq:lep}).
This is a particular case of 
the Cabibbo--Gatto relation 
for a resonant, spin-one interaction \cite{CG}, valid for any
hadronic final state. Thus, an analogous ``Cabibbo-Gatto'' 
formula exists for $e^+e^-\ra \omega \ra \pi^+\pi^0\pi^-$. In this
manner, $\Gamma(\omega\ra e^+e^-)$ and 
$f_{\omega_I \gamma}$, via Eq.~(\ref{eq:lep}), can both be 
inferred from the \ethrpi data~\cite{3pi}. We use 
$\Gamma({\omega \ra e^+e^-})=0.60\pm.02$ keV~\cite{PDG} in what follows.

We can now calculate $\Gamma({\rho^0\ra e^+e^-})$ and hence
$f_{\rho_I\gamma}$.
Recalling 
Eq.~(\ref{ff-isofit}) we find
\be
\sigma(e^+e^-\ra\rho_I\ra\pi^+\pi^-)
=\frac{\pi\alpha^2}{3}\frac{(s-4m_\pi^2)^{3/2}}{s^{5/2}}
|F_\rho(s)|^2 \Bigg|_{s=m_\rho^2} \;,
\label{rhoIcross}
\ee
which when combined with
Eq.~(\ref{picase}) 
yields
\be
\Gamma(\rho_I^0\ra e^+e^-)=\frac{\alpha^2}{36}
\frac{(m_\rho^2-4m_\pi^2)^{3/2}}{m_\rho^3}|F_\rho(m_\rho^2)|^2\Gamma_\rho \;,
\label{Gree}
\ee  
where $\Gamma_\rho=\Gamma(\rho\ra\pi^+\pi^-)$,
allowing us to determine $f_{\rho_I\gamma}$
from Eq.~(\ref{eq:lep}).

We note in passing that it
is quite common in the literature to see the $\omega$ contribution
to the pion form-factor
expressed in terms of $\omega$ partial widths \cite{Bark,Ben}. Such
an expression follows 
from our Eq.~(\ref{ff-geff1}), in concert with Eqs.~(\ref{eq:gpp}) 
and (\ref{eq:lep}), to yield
\bea
F_\pi(s)
=F_\rho(s)
+ \sqrt{\frac{36\Gamma({\omega\ra e^+e^-})\Gamma({\omega\ra\pi^+\pi^-})}{
m_\omega^2\alpha^2\beta_\omega^3}}\frac{m_\omega^2}{s-
m_\omega+im_\omega\Gamma_\omega},
\label{eq:rit}
\eea
where $\beta_\omega=(1-4m_\pi^2/m_\omega^2)^{1/2}$ and
we replace 
$f_{\rho_I\gamma}g_{\rho_I\pi\pi}/(s-m_\rho+im_\rho\Gamma_\rho)$ with
$F_\rho(s)$ as earlier. 
Thus, our Eq.~(\ref{Br_final}) is explicitly
equivalent to the determinations of Br($\omega\ra\pi^+\pi^-$) found
in Refs.~\cite{Bark,Ben}. 

\section{Results and Discussion}

We now use our fits of Ref.~\cite{GO} to compute 
Br($\omega\ra\pi^+\pi^-$), $\Gamma(\rho\ra e^+e^-)$, and other
associated parameters, in addition to their errors. 
Our fits to the pion form-factor 
data~\cite{wd85}, noting Eq.~(\ref{ff-isofit}), adopt parametrizations of
$F_\rho(s)$ consistent with the following theoretical constraints.
That is, analyticity requires that 
$F_\rho(s)$ be real below threshold,
$s=4 m_\pi^2$, charge conservation requires $F_\rho(0)=1$, and
unitarity and time-reversal invariance 
requires its phase be that of $l=1$, $I=1$ $\pi$-$\pi$
scattering for $s$ where the latter is elastic~\cite{HL}. 
For the present work we shall concentrate on four of
these choices for $F_\rho(s)$, labeled, as per Ref.~\cite{GO}, 
A, B, C, and D, in which
$\prw$
is an explicit fitting parameter. These four 
fits assume $\prw$ to be 
a real constant in the resonance region, for the current \etwopi data 
supports neither a phase nor $s$-dependent pieces~\cite{GO}.

Table \ref{one} shows our results for $\Gamma({\rho^0\ra e^+e^-})$ 
and Br$(\rho^0\ra e^+e^-)\equiv\Gamma({\rho^0\ra e^+e^-})/\Gamma_\rho$. 
as determined
from Eq.~(\ref{Gree}).
We find the following average values:
\be
\Gamma({\rho^0\ra e^+e^-})
\!=\!7.11\pm 0.08\pm 0.25\;{\rm keV},\, {\rm Br}(\rho^0\ra e^+e^-)\!=\!
(4.63\pm 0.05 \pm 0.07)\times10^{-5},
\ee
where the second error on $\Gamma({\rho^0\ra e^+e^-})$ is 
the theoretical systematic error associated with
model choice~\cite{syserr}, and all other errors are statistical.
$\Gamma_\rho$ from Fit D is significantly lower
than those from the other fits and leads to a significantly lower value for
$\Gamma({\rho^0\ra e^+e^-})$, indeed one commensurate with the value  of 
$6.77\pm 0.10 \pm 0.30$  keV
reported in Ref.~\cite{Bark}. This is likely consequent to
the choice of the Gounaris--Sakurai form factor \cite{GS} in both fits; our
other fits use a Heyn--Lang form factor \cite{HL}.
Such model
dependence also plagues the extraction of the $\rho$ parameters 
$m_\rho$ and $\Gamma_\rho$, as discussed following Eq.~(\ref{pole}). 

Using $\Gamma(\rho_I\ra e^+ e^-)$ of Table \ref{one} and Eq.~(\ref{eq:lep})
yields $f_{\rho_I\gamma}$ and $r_\gamma$, using
 $f_{\omega_I\gamma}$ computed from $\Gamma(\omega\ra e^+e^-)$ of 
Ref.~\cite{PDG}.
In the SU(6) limit $r_\gamma$ is 3; 
the ``finite width'' correction~\cite{GS,renard}, as 
seen in Table \ref{two}, is $\sim10$\%, as also found in Ref.~\cite{klingl}, 
and hence significant. 
Including this correction as per Eq.~(\ref{replace}) 
gives us perhaps a 
more realistic value of 
$\prw(m_\omega^2)$~\cite{footnote}, 
and its model dependence appears to be modest, allowing
us to determine an average value of 
\be
\prw(m_\omega^2)=-3900\pm 300\,{\rm MeV}^2\;,
\label{newpirw}
\ee
again some 10\% larger than our value of 
$\prw(m_\omega^2)=-3500\pm 300\,{\rm MeV}^2$ in Ref.~\cite{GO}
using $r_\gamma=3$.

Our preferred determination of 
Br$(\omega\ra\pi^+\pi^-)$, Eq.~(\ref{Br_final}), does not require
$r_\gamma$, and we find 
\be
{\rm Br}(\omega\ra\pi^+\pi^-)=1.9\pm0.3\%\;.
\ee
Barkov {\it et al.}, noting Eq.~(\ref{eq:rit}) and the discussion
thereafter, obtain Br$(\omega\ra\pi\pi)=2.3\pm0.4\%$~\cite{Bark} 
with the same data set~\cite{wd85} used here. 
%!
We agree closely, however, with 
the result of Bernicha {\it et al.}, $1.85\pm0.30\%$~\cite{BCP}, obtained 
from the same data~\cite{wd85}. 
Their relation for the branching ratio, 
Eq.(42)~\cite{BCP}, is our Eq.~(\ref{Br2}), though they use 
the parameters ${\overline m}_\rho$ and ${\overline \Gamma}_\rho$,
noting Eq.~(\ref{pole}), in place of $m_\rho$ and $\Gamma_\rho$ and
use $\Gamma_{\rho\ra e^+e^-} = 6.77$ keV to compute the leptonic
coupling $f_{\rho_I\gamma}$~\cite{BCP}. The latter effects compensate, so that
we would expect to find a branching ratio comparable to theirs. 
The data set we have adopted~\cite{wd85}
contains 30 data points for center of mass energies between
750 and 810 MeV, the region likely most relevant for the determination of
$\Gamma(\omega\ra\pi^+\pi^-)$. The older work of 
Benaksas {\it et al.}~\cite{Ben} and
Quenzer {\it et al.} \cite{Q} find 
Br$(\omega\ra\pi\pi)=3.6\pm0.4\%$ and 
Br$(\omega\ra\pi\pi)=1.6\pm0.9\%$, respectively, though both experiments
possess less than 10 data points in the energy region of interest.

We can also compute Br($\omega\ra\pi^+\pi^-$) using
Eqs.~(\ref{Br2}) or (\ref{Br3}) and (\ref{replace}), 
as shown in Table \ref{three}. 
Apparently it makes little difference whether we use
Eq.~(\ref{Br_final}) or Eq.~(\ref{Br2}), though the former, 
our preferred analysis, possesses essentially no parametrization
dependence. Br$^{(3)}(\omega\ra\pi\pi)$, from Eq.~(\ref{Br3}), is
substantially larger, though this may be an artifact of using
the true S-matrix pole position $z_\rho$ in Eq.~(\ref{Br3}). 
If we were to replace $z_\rho$ with
$m_\rho^2 - i m_\rho \Gamma_\rho$, noting the discussion surrounding
Eq.~(\ref{pole}), the values, as shown in parentheses,
%!
would differ less, even though we were obliged to 
assume that
$\overline{\Pi}_{\rho\omega}(m_\omega^2)$ and
$\widetilde{\Pi}_{\rho\omega}(m_\omega^2)$ are the same. 

In summary, we have elucidated the connection between 
$\prw(m_\omega^2)$ and Br($\omega\ra\pi^+\pi^-$) and shown 
how different methods of determining 
Br($\omega\ra\pi^+\pi^-$) would be equivalent
were it possible to evaluate $\prw(z_\omega)$. In practice, the
methods are different, yet, nevertheless,
it seems that a plurality of methods of computing
Br($\omega\ra\pi^+\pi^-$) yield roughly comparable results.

\acknowledgments
This work was supported by the U.S. Department of Energy 
under grant \# DE--FG02--96ER40989.

\nc{\m}{$m_\rho$ (MeV)}
\nc{\G}{$\Gamma_\rho$ (MeV)}
\nc{\gree}{$\Gamma({\rho\ra e^+ e^-})$ (keV)}
\nc{\Bree}{Br$({\rho\ra e^+e^-})\times (10^{5})$}
\nc{\frg}{$f_{\rho_I\gamma}$ (GeV$^2$)}
\nc{\zm}{$\overline{m}_\rho$ (MeV)}
\nc{\zg}{$\overline{\Gamma}_\rho$ (MeV)}
\begin{table}[htb]
\caption{
The results of our fits~\protect{\cite{GO}} to the pion form-factor 
and the corresponding values of
$\Gamma(\rho^0\ra e^+e^-)$, noting Eq.~(\protect{\ref{Gree}}), and 
Br$(\rho^0\ra e^+e^-)$.
Also shown are the $\rho$ parameters, $\overline{m}_\rho$
and $\overline{\Gamma}_\rho$,
defined from the pole position $z_\rho$, as in Eq.~(\protect{\ref{pole}}).
}
\begin{tabular}{ccccccc}
Fit & \m\cite{GO} & \G\cite{GO} &\gree & \Bree & \zm & \zg          \\
\hline
A&$763.1\pm3.9$&$153.8\pm1.2$&$7.27\pm0.08$&$4.73\pm0.05$&
        $756.3\pm1.2$ & $141.9\pm3.1$\\
B&$771.3\pm1.3$&$156.2\pm0.4$&$7.24\pm0.08$&$4.63\pm0.06$&
        $757.0\pm1.0$ & $141.7\pm3.0$\\
C&$773.9\pm1.2$&$157.0\pm0.4$&$7.19\pm0.08$&$4.58\pm0.05$&
        $757.0\pm1.0$ & $141.7\pm3.0$\\
D&$773.9\pm1.2$&$146.9\pm3.4$&$6.73\pm0.10$&$4.58\pm0.05$&
        $757.0\pm1.0$ & $141.7\pm3.0$\\
\end{tabular}
\label{one}
\end{table}

\vspace{2cm}

\nc{\prwm}{$\widetilde{\Pi}_{\rho\omega}(m_\omega^2)$}
\nc{\ff}{$f_{\rho_I\gamma}/f_{\omega_I\gamma}$}
\nc{\B}{Br$(\omega\ra\pi^+\pi^-)$ }
\nc{\BB}{Br$^{(2)}(\omega\ra\pi^+\pi^-)$ }
\nc{\BBB}{Br$^{(3)}(\omega\ra\pi^+\pi^-)$ }
\begin{table}[htb]
\caption{
Results for the effective \rhoom mixing element, $\prw$,
and the branching ratio \B, from Eq.~(\protect{\ref{Br_final}}), using the
fits of Ref.~\protect{\cite{GO}}. 
$f_{\omega_I\gamma}$ follows from
Eq.~(\protect{\ref{eq:lep}}) 
and the parameters of 
Ref.~\protect{\cite{PDG}}. We also show the value of $\prw$ 
which results from using Eq.~(\protect{\ref{replace}}) with 
$f_{\rho_I\gamma}/f_{\omega_I\gamma}$, as per 
Eqs.~(\protect{\ref{eq:lep}}) and (\protect{\ref{picase}}), again
using the fits of Ref.~\protect{\cite{GO}}. 
}
\begin{tabular}{cccccc}
Fit&\prwm (MeV$^2$)\cite{GO}&\frg  &  \ff  &\prwm (MeV$^2$) &\B\\    
\hline
A&$-3460\pm290$&$0.120\pm0.001$&$3.36\pm0.07$&$-3870\pm320$&$1.87\pm0.30\%$\\
B&$-3460\pm290$&$0.122\pm0.001$&$3.40\pm0.06$&$-3920\pm330$&$1.87\pm0.30\%$\\
C&$-3460\pm290$&$0.122\pm0.001$&$3.41\pm0.06$&$-3930\pm330$&$1.87\pm0.30\%$\\
D&$-3460\pm290$&$0.118\pm0.001$&$3.30\pm0.06$&$-3800\pm330$&$1.87\pm0.30\%$\\
\end{tabular}
\label{two}
\end{table}

\vspace{2cm}

\begin{table}[htb]
\caption{
The branching ratio \B from 
our preferred method, Eq.~(\protect{\ref{Br_final}}), 
compared with the alternatives \BB,
Eq.~(\protect{\ref{Br2}}), and \BBB, Eq.~(\protect{\ref{Br3}}). 
In parentheses
we give the values for the branching ratio as determined by 
Eq.~(\protect{\ref{Br3}})
but replace $z_\rho$ with $m_\rho^2-im_\rho\Gamma_\rho$, noting
the discussion preceding Eq.~(\protect{\ref{pole}})
and the results of Table~\protect{\ref{one}}.
}
\begin{tabular}{ccccc}
Fit & \B          & \BB          & \BBB\\
\hline
A&$1.87\pm0.30\%$&$1.93\pm0.32\%$&$2.41\pm0.39\%\,\,(2.15\pm0.35\%$)\\
B&$1.87\pm0.30\%$&$1.97\pm0.32\%$&$2.50\pm0.40\%\,\,(2.19\pm0.35\%$) \\
C&$1.87\pm0.30\%$&$1.96\pm0.32\%$&$2.51\pm0.40\%\,\,(2.19\pm0.35\%$) \\
D&$1.87\pm0.30\%$&$1.95\pm0.32\%$&$2.20\pm0.37\%\,\,(2.20\pm0.35\%$) \\
\end{tabular}
\label{three}
\end{table}

\end{document}